\documentclass[preprint]{aastex}
\usepackage{natbib}
\usepackage{booktabs}
\usepackage{threeparttable}

\usepackage{longtable}
\usepackage{tabularx}
\usepackage{lineno}

\newcommand{\fexxi}{Fe \scriptsize{XXI} \normalsize}
\newcommand{\fexviii}{Fe \scriptsize{XVIII} \normalsize}
\newcommand{\fexvi}{Fe \scriptsize{XVI} \normalsize}
\newcommand{\cii}{C \scriptsize{II} \normalsize}
\newcommand{\siiv}{Si \scriptsize{IV} \normalsize}

\begin{document}

\title{Simultaneous Observations of Chromospheric Evaporation and Condensation during a C-class Flare}

\author{Dong~Li\altaffilmark{1,2,3}, Zhenxiang~Hong\altaffilmark{1,2}, and Zongjun~Ning\altaffilmark{1,2}}

\affil{$^1$Key Laboratory of Dark Matter and Space Astronomy, Purple Mountain Observatory, CAS, Nanjing 210023, PR China \\
     $^2$School of Astronomy and Space Science, University of Science and Technology of China, Hefei 230026, PR China \\
     $^3$CAS Key Laboratory of Solar Activity, National Astronomical Observatories, Beijing 100101, PR China}
     \altaffiltext{}{Correspondence should be sent to: lidong@pmo.ac.cn}

\begin{abstract}
We explored simultaneous observations of chromospheric evaporation
and condensation during the impulsive phase of a C6.7 flare on 9 May
2019. The solar flare was simultaneously observed by multiple
instruments, i.e., the New Vacuum Solar Telescope (NVST), the
Interface Region Imaging Spectrograph, the Atmospheric Imaging
Assembly (AIA), the Fermi, the Mingantu Spectral Radioheliograph,
and the Nobeyama Radio Polarimeters. Using the single Gaussian
fitting and the moment analysis technique, red-shifted velocities at
slow speeds of 15$-$19~km~s$^{-1}$ are found in the cool lines of
\cii\ and \siiv\ at one flare footpoint location. Red shifts are
also seen in the H$\alpha$ line-of-sight (LOS) velocity image
measured by the NVST at double footpoints. Those red shifts with
slow speeds can be regarded as the low-velocity downflows driven by
the chromospheric condensation. Meanwhile, the converging motions
from double footpoints to the loop top are found in the
high-temperature EUV images, such as AIA~131~{\AA}, 94~{\AA}, and
335~{\AA}. Their apparent speeds are estimated to be roughly
126$-$210~km~s$^{-1}$, which could be regarded as the high-velocity
upflows caused by the chromospheric evaporation. The nonthermal
energy flux is estimated to be about 5.7$\times$10$^{10}$
erg~s$^{-1}$~cm$^{-2}$. The characteristic timescale is roughly
equal to 1~minute. All these observational results suggest an
explosive chromospheric evaporation during the flare impulsive
phase. While a HXR/microwave pulse and a type III radio burst are
found simultaneously, indicating that the explosive chromospheric
evaporation is driven by the nonthermal electron.
\end{abstract}


\section{Introduction}
Solar flare generally shows a sudden brightening feature on the Sun,
which is often accompanied by a quick and violent energy release via
magnetic reconnection
\citep[e.g.,][]{Fletcher11,Benz17,Chen20,Tan20}. The wavelength
range of flare radiation is quite broad, i.e., from the
radio/microwave through the white light and extreme ultraviolet
(UV/EUV) to soft/hard X-rays (SXR/HXR) and even to the $\gamma$-ray
\citep[e.g.,][]{Masuda94,Su13,Dominique18,Yan18,Yan18b,Lysenko20,Li21}.
In the standard model of a solar flare such as the CSHKP
flare model \citep{Carmichael64,Sturrock66, Hirayama74,Kopp76}, a
huge amount of magnetic energies are released via reconnection in
the local corona \citep{Shibata11,Li16,Jiang21}, i.e., as described
in the two-dimensional reconnection model
\citep{Sturrock64,Priest02,Lin05}. In the reconnection site, the
local plasma will be heated to the thermal energy, while the
electron will be accelerated to the nonthermal energy.
Subsequently, some accelerated electrons could travel upward
along the open magnetic lines and escape away or may propagate
upward along the closed magnetic field lines, while the other
electron beams will transport downward along the closed magnetic
lines, and then precipitate in the chromosphere. This has been
known as `bi-directional outflows'
\citep[e.g.,][]{Innes97,Mann11,Liu13,Lid19,Warmuth20,Yan21a}. In the
chromosphere, the precipitated electrons could heat the localized
plasma rapidly up to a very high temperature via coulomb collisions,
resulting overpressure in the localized chromosphere. The
overpressure can drive the chromospheric material upward along the
new formed flare loop into the overlying corona at a fast speed, for
instance, an order magnitude of hundreds of km~s$^{-1}$. The hot
material fills up the newly flare loop rapidly in a process regarded
as `chromospheric evaporation'
\citep[e.g.,][]{Fisher85,Teriaca03,Brosius15,Dudik16,Tian18}. In
this process, double footpoints at the loop legs are often
generated in the HXR or microwave emissions, and the flare ribbons
could be seen in H$\alpha$, white light, UV or low-temperature EUV
wavelengths, while flare loops can be found in the SXR and
high-temperature EUV images
\citep[e.g.,][]{Sui03,Asai06,Temmer07,Krucker08,Brosius16,Chen17,Song18}.
Due to the momentum conservation, the overpressure of the
chromosphere can simultaneously cause downward mass motions into the
underlying chromosphere at a slow speed, a process termed
`chromospheric condensation'
\citep{Kamio05,Zhang16a,Libbrecht19,Graham20,Yu20}.

Observations of the chromospheric evaporation have been widely
reported in the flare radiation at wavelengths of radio, UV/EUV, and
HXR. In the radio dynamic spectrum, the higher frequency suddenly
being suppressed and drifting to lower frequencies can be regarded
as the radio evidence of chromospheric evaporation
\citep{Aschwanden95,Karlicky98,Ning09}. In the HXR images, the HXR
emissions often appear firstly at two footpoint sources, and then
rise upward along their loop legs, subsequently merge into a single
source at the loop-top region
\citep{Liu06,Liu08,Ning10,Ning11a,Ning11}. The same observational
features can also be seen in the high-temperature EUV images, for
instance, the drastic mass motions from the double footpoints along
the hot flare loop to the loop top \citep{Li17,Zhang19}. This is
regarded as the HXR/EUV signature of chromospheric evaporation
\citep[e.g.,][]{Liu06,Ning11,Zhang19}. The mergence/move speed could
be as high as about 200~km~s$^{-1}$, which is roughly equal to the
upflow speed of the hot evaporated materials. Thanks to the
spectroscopic observations, the chromospheric evaporation in solar
flares can be well diagnosed by the line profile
\citep[e.g.,][]{Milligan09,Doschek13,Tian14,Li15,Polito17,Dep21}. In
the typical chromospheric evaporation, the hot coronal lines always
appear high-velocity blue shifts, which are attributed to the fast
upward mass motions. While the cool chromospheric or transition
region lines could exhibit low-velocity red shifts due to the slow
downward mass motions, or may show blue shifts without associated
clear downflows. The former one is often regarded as `explosive
evaporation' \citep{Milligan06a,Zhang16b,Brosius17,Li17}, and the
latter one is referred as `gentle evaporation'
\citep{Milligan06b,Sadykov15,Li19}. These two types of chromospheric
evaporation are often depending on the heating energy of electron
beams, for instance, a critical threshold of about
10$^{10}$~erg~s$^{-1}$~cm$^{-2}$
\citep[e.g.,][]{Fisher85,Kleint16,Sadykov19}. However, the threshold
value of the deposited energy flux is not a constant, it is
sensitive to the duration of the heating and the location of the
energy deposition \citep{Reep15,Polito18}. \cite{Rubio15} stated
that the explosive evaporation could also be caused by stochastic
accelerated electrons with a low energy flux that is less than
10$^{10}$~erg~s$^{-1}$~cm$^{-2}$.

Chromospheric evaporation has been detected in a large number of
solar flares
\citep[e.g.,][]{Ding96,Li04,Falewicz09,Raftery09,Chen10,Veronig10,Li11,Brosius13,Young15,Brosius18,Tian18},
largely benefiting from the spectroscopic observations, such as, the
Bragg Crystal Spectrometer on board Yohkoh, the Coronal Diagnostic
Spectrometer aboard the Solar and Heliospheric Observatory,
the Extreme-ultraviolet Imaging Spectrometer experiment aboard
Hinode, the Interface Region Imaging Spectrograph (IRIS), et cetera.
They can be detected in the pre-flare phase \citep{Brosius10,Li18},
impulsive phase \citep{Graham15,Li17,Sadykov19} and gradual/decay
phase of solar flares \citep{Czaykowska99,Li12,Brannon16}. The
chromospheric evaporation could be explained by the electron driven
\citep{Tian15,Warren16,Lee17} or the thermal conduction driven
\citep{Czaykowska01,Battaglia15,Ashfield21}, as well as the
dissipation of Alv\'{e}n waves \citep{Fletcher08,Reep16}. The first
one states that the nonthermal energy produced by electron beams
plays an important role in the evaporation process, while the second
one emphasizes that the evaporation motion is directly driven by the
thermal energy. In this paper, we present observational evidences of
an explosive chromospheric evaporation during the impulsive phase of
a solar flare, and it appears to be driven by the nonthermal
electron.

\section{Observations and Instruments}
A solar flare occurred on 9 May 2019 in active region (AR) NOAA
12740, which was close to the solar disk center, e.g., N09E13. Based
on the solar
monitor\footnote{https://www.lmsal.com/solarsoft/latest\_events\_archive/events\_summary/2019/05/09/gev\_20190509\_0540/index.html},
the flare was a C6.7 class, and it started at about 05:40~UT,
reached its maximum at around 05:51~UT. It was simultaneously
measured by the IRIS \citep{Dep14}, the New Vacuum Solar Telescope
\citep[NVST,][]{Liu14}, the Atmospheric Imaging Assembly
\citep[AIA,][]{Lemen12} and the Helioseismic and Magnetic Imager
\citep[HMI,][]{Schou12} for the Solar Dynamics Observatory
\citep[SDO,][]{Pesnell12}, the X-ray Telescope
\cite[XRT,][]{Golub07} of the Hinode mission, as well as the
Geostationary Operational Environmental Satellite
\citep[GOES-16,][]{Lotoaniu17}, the Fermi Gamma Ray Burst Monitor
\citep[GBM,][]{Meegan09}, the Mingantu Spectral Radioheliograph
\citep[MUSER,][]{Yan21}, and the Nobeyama Radio Polarimeters
\citep[NoRP,][]{Nakajima85}, as listed in table~\ref{tab1}.

The NVST is a one meter aperture vacuum telescope located at Fuxian
Solar Observatory, which is operated by Yunnan Observatories. It can
provide high-resolution images of the chromosphere and photosphere,
corresponding to the channels of H$\alpha$ and TiO, respectively
\citep{Liu14,Yan20a}. In this paper, the NVST H$\alpha$ images at
the line core (6562.8~{\AA}) and two off bands ($\pm$0.5~{\AA}) are
acquired from 05:35~UT to 09:58~UT on 9 May 2019. Using the speckle
masking method \citep{Weigelt77,Lohmann83,Xiang16}, these images are
reconstructed from the Level 1 to Level 1+, which can be accessed at
the NVST
website\footnote{http://fso.ynao.ac.cn/cn/datashow.aspx?id=2344}.
They have a spatial pixel size of $\sim$0.165\arcsec\ and a time
cadence of $\sim$43~s.

The IRIS provides multi-channel UV imaging spectrograph in space,
which is mainly focused on the solar chromosphere and transition
region \citep{Dep14}. On 9 May 2019, IRIS scanned the AR NOAA 12740 in
a `very large dense 320-step raster' mode, which had a max
field-of-view (FOV) of $\sim$279\arcsec~$\times$175\arcsec. The step
size and cadence were 0.35\arcsec\ and $\sim$16.2~s, respectively,
while the pixel scale along the slit was $\sim$0.33\arcsec. The far
UV (FUV) spectra and images are acquired from the IRIS level 2 data,
which has been pre-processed and calibrated by the IRIS
team. Here, we used the spectral lines at spectral windows of
`\siiv~1394~{\AA}', `\siiv~1403~{\AA}', and `\cii~1336~{\AA}', as
well as the Slit-Jaw Imager (SJI) at 1400~{\AA}. The FUV spectra has
a spectral dispersion of about 0.0256~{\AA}~pixel$^{-1}$, and the
SJI 1400~{\AA} images had a time cadence of $\sim$65~s.

The SDO/AIA is designed to take full-disk solar images in multiple
wavelengths nearly simultaneously \citep{Lemen12}, and the SDO/HMI
is designed to investigate the solar magnetic field and oscillatory
features on the Sun \citep{Schou12}. Here, the AIA images at
high-temperature EUV wavelengths of 131~{\AA} ($\sim$10~MK),
94~{\AA} ($\sim$6~MK), and 335~{\AA} ($\sim$3~MK) are used, which
have a time cadence of 12~s. We also used the HMI line-of-sight
(LOS) magnetogram to show the distribution of magnetic fields. Both
the AIA EUV images and the HMI LOS magnetogram have been
pre-processed via the AIA and HMI standard routines, such as
`aia\_prep.pro' and `hmi\_prep.pro'. Then, they both have a spatial
scale of 0.6\arcsec~pixel$^{-1}$. The Hinode/XRT provides solar
images via nine X-ray filters, which measures the coronal plasma
from about 1~MK to approximately 20~MK, and they were calibrated
using the standard XRT program such as `xrt\_prep.pro' before
analyzing \citep{Golub07}. During the C6.7 flare, the XRT takes
synoptic data in the form of full-disk images with both long and
short exposures from 05:40:03~UT to 05:43:49~UT, and the spatial
scale is $\sim$4.1\arcsec~pixel$^{-1}$.

The C6.7 flare was also recorded by the MUSER, NoRP, Fermi/GBM, and
GOES at radio, HXR and SXR emissions. The MUSER could provide the
radio dynamic spectrum between 0.4~GHz and 2.0~GHz. It has forty
4.5-m reflector antennas distributed in three spiral arms, which can
measure dual circular polarizations with a time resolution of
0.025~s \citep{Yan09,Yan21}. The NoRP and the Fermi/GBM observed the
entire Sun with multiple channels in microwave and X-ray ranges,
respectively. In this study, we used the radio dynamic spectrum
detected by the No.33 antenna of MUSER, because we did not find any
pronounced polarization in the radio bursts for the forty antennas
during the observed time. The microwave flux at the frequency of
3.75~GHz with a time cadence of 1~s was used \citep{Nakajima85}. The
HXR flux at 26.6$-$50.4~keV recorded by the n3 detector was used,
which has a time resolution of about 4.096~s \citep{Meegan09}.
Noting that the HXR light curve was derived from the cspec file
rather than the ctime file, because we also perform the X-ray
spectral analysis to look the deposited energy flux. The SXR flux at
1$-$8~{\AA} recorded by the GOES-16 was also used, which has a time
cadence of 1~s.

\section{Data analysis and Results}
Figure~\ref{flux} presents the overview of the solar flare on 9 May
2019. Panel~(a) plots the SXR~1$-$8~{\AA} light curve
recorded by the GOES-16, which suggests that the flare peak flux
could exceed $10^{-5}$~Watts~m$^{-2}$. It should be pointed
out that the GOES XRS flux has been corrected by a constant factor
of about 1.5 for above the C1-class flare \citep{Woods17}. However,
it was regarded as a C6.7 class flare in the solar monitor, which
was from the GOES-14 SXR flux. In this paper, we also regard it as a
C6.7 class, which is similar to the previous study \citep{Yan20b}.
It began at about 05:40~UT, and peaked at roughly 05:51~UT. The HXR
and microwave fluxes recorded by the Fermi/GBM and NoRP are also
presented in Figure~\ref{flux}~(a). They both show three main pulses
(`1'$-$`3') during $\sim$05:43$-$05:49~UT, which is in the impulsive
phase of the C6.7 flare and could be regarded as the quasi-periodic
pulsations \citep{Hong21}. In panel~(b), three groups of transient
bursts can be seen in the radio dynamic spectrum measured by the
MUSER during the same time interval, and they drift rapidly from the
high frequency to the low frequency over a very short time. Thus,
they could be regarded as three groups of type III radio bursts,
revealing a well one-to-one correspondence with the three
HXR/microwave pulses, as indicated by the Roman numerals in
Figure~\ref{flux}~(a) and (b). All these observations imply that
pulsed nonthermal electrons are intermittently accelerated by the
repeated magnetic recognition during the flare impulsive phase.

In Figure~\ref{image}, we show the multi-wavelength snapshots of the
C6.7 flare with the same FOV of about 130\arcsec$\times$130\arcsec.
Panels (a) and (b) present high-temperature coronal images at nearly
the same time during the solar flare, i.e., XRT Al\_mesh, and
AIA~131~{\AA}. They both show a bright loop-like structure, as
outlined by the purple contour. Thus, the XRT and AIA images can be
co-aligned with the cross-correlation method. The loop-like
structure observed in the high-temperature coronal images could be
regarded as a potential candidate of the hot flare loop, as
indicated by the purple arrow. Here, the XRT Al\_mesh image is
selected because it corresponds pixels in the short exposure that
are not saturated. Panels~(c) \& (d) present the SJI~1400~{\AA} and
H$\alpha$~6562.8~{\AA} images measured by the IRIS and NVST. The
bright and strip features can be simultaneously seen in the
SJI~1400~{\AA} and H$\alpha$ line-core images, which might be
regarded as the flare ribbons, as indicated by the cyan arrows.
Moreover, one of the flare ribbons was crossed by the IRIS slit, as
indicated by the dashed vertical line in panel~(c). Panels~(e) \&
(f) plot the NVST images in two H$\alpha$ off bands at
$\pm$0.5~{\AA}, and they are used to obtain the LOS velocity image.
The NVST and IRIS data can be co-aligned by using the
cross-correlation technique, because they both show the bright flare
ribbons. While the AIA~1600~{\AA} image (panel~g) is applied to
co-align with the SJI~1400~{\AA} image, since they both contain the
continuum emissions from the temperature minimum
\citep[see,][]{Li15,Tian15}. Panels~(h) \& (i) show the AIA~94~{\AA}
running difference map and the HMI LOS magnetogram. A bright
loop-like structure appears in the AIA~94~{\AA} running difference
map. To clearly see the bright loop, we plot the AIA~94~{\AA}
running difference map with a small FOV of about
30\arcsec$\times$30\arcsec, as shown by the zoomed image in
panel~(h). It is very similar to the high-temperature images in
AIA~131~{\AA} and XRT/Al\_mesh. Moreover, the two-end locations of
the loop-like structure are rooted in the positive (Ft1) and
negative (Ft2) magnetic fields, respectively, as indicated by the
cyan curve in panel~(i). In panel~(h), the overplotted contours
represent the AIA~94~{\AA} emissions at three fixed instances of
time. They begin to brighten in two separated locations, for
instance, the two red contours at about 05:43:35~UT. Then they move
closely along the loop-like structure, as indicated by the two blue
contours at about 05:43:59~UT. At about 05:44:35~UT, only one bright
source (yellow) can be seen, which can be regarded as the loop-top
source. All these observational facts confirm the bright loop-like
structure could be regarded as the hot flare loop, and the two
brightening locations can be regarded as the double footpoints (Ft1
and Ft2) connected by the flare loop. \cite{Yan20b} suggested that
the AR NOAA 12740 was made up of a large compact leading negative
polarity and some small diffuse positive following polarities. The
C6.7 flare took place at the northeast edge of the leading negative
magnetic field, as shown in panels~(c)$-$(i). An AR filament located
at the northeast of the leading sunspot can be seen in the H$\alpha$
image (panel~d), and it was a failed filament eruption \citep[see
detail,][]{Yan20b}.

\begin{figure}
\centering
\includegraphics[width=\linewidth,clip=]{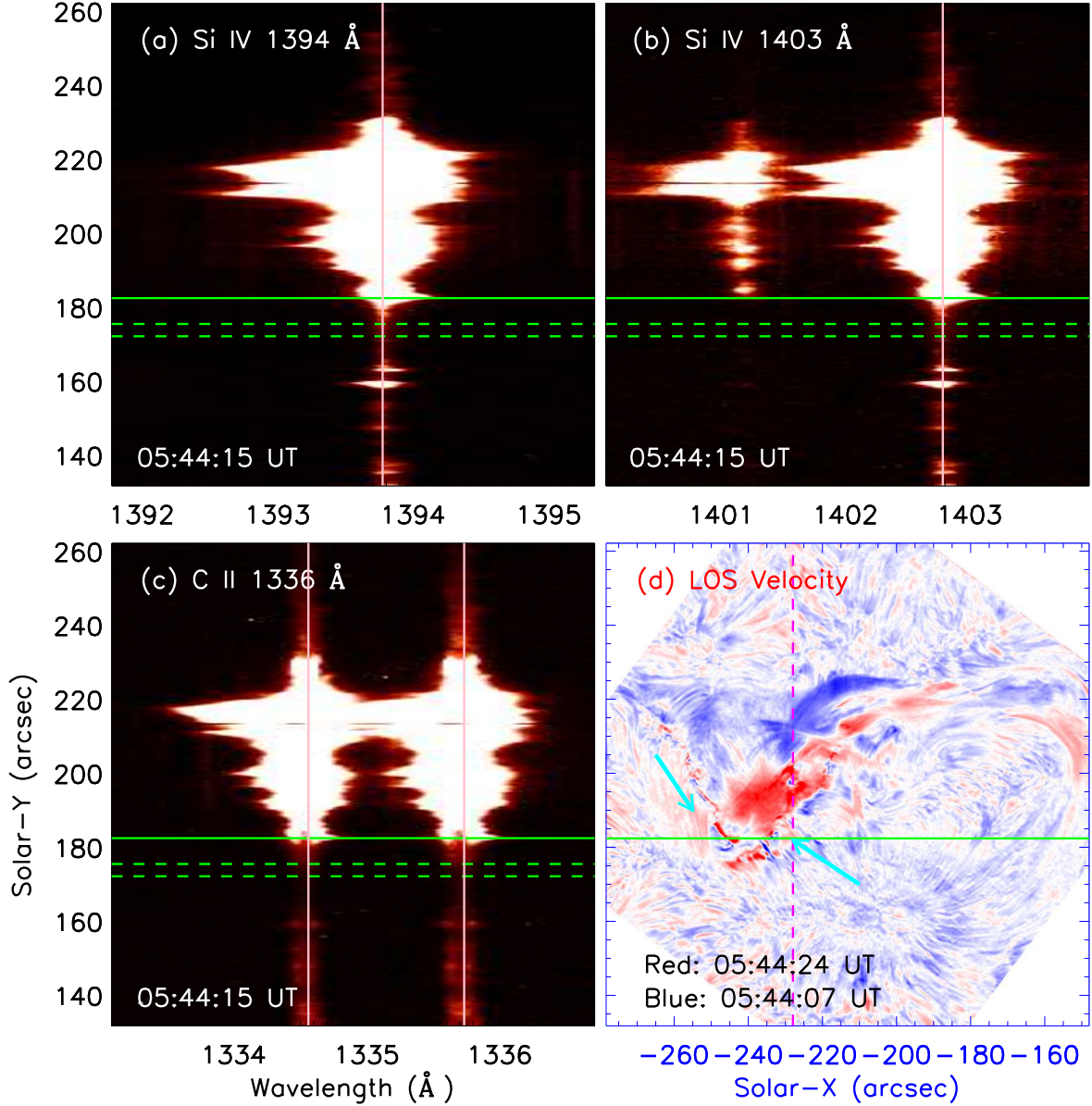}
\caption{Panels~(a)$-$(c): IRIS spectra at three FUV windows of
`\siiv~1394~{\AA}' (a), `\siiv~1403~{\AA}' (b), and
`\cii~1336~{\AA}' (c). The horizontal lines mark the flare footpoint
location (solid) and quiet region (dashed). The vertical
pink lines mark the positions of the reference line centers.
Panel~(d): LOS velocity image derived from two H$\alpha$ off bands
at $\pm$0.5~{\AA}, measured by the NVST. The vertical magenta line
represents the IRIS slit, and the cyan arrows indicated two flare
footpoints. \label{spect}}
\end{figure}

One slit of the IRIS crossed the flare ribbon, which is well used to
study the chromospheric evaporation
\citep[e.g.,][]{Li15,Tian15,Young15,Polito17,Li18,Graham20}. We then
plot the IRIS spectra at three FUV windows of `\siiv~1394~{\AA}',
`\siiv~1403~{\AA}', and `\cii~1336~{\AA}', as shown in
Figure~\ref{spect}~(a)$-$(c). Panel~(d) draws the LOS velocity image
obtained by the two off-bands ($\pm$0.5~{\AA}) H$\alpha$ images at
almost the same time. Thus, it is only a qualitative description
rather than the quantitative calculation. The interested site is
marked by the solid green line, which is regarded as one footpoint
(Ft2) of the hot flare loop, and it appears a weak redshifted
velocity. The other footpoint (Ft1) also exhibits weak redshifted
velocities in the H$\alpha$ LOS velocity image, as indicated by the
cyan arrow. But we cannot check the IRIS spectra at the other
footpoint, because it was not crossed by the IRIS slit during the
C6.7 flare. We notice that both IRIS spectral lines and H$\alpha$
LOS velocity image shown strong blueshifted velocity at
Y$\sim$200$-$220\arcsec\ along the IRIS slit, and then become strong
redshifted velocity at Y$\sim$180$-$200\arcsec. However, they are
out of the scope of this study., because they are far away from the
hot flare loop, as shown in Figure~\ref{image}. The line spectra at
the quiet region (outlined by the dashed green lines) are also
selected to determine their rest wavelengths or reference line
centers, as indicated by the vertical pink lines.

\begin{figure}
\centering
\includegraphics[width=\linewidth,clip=]{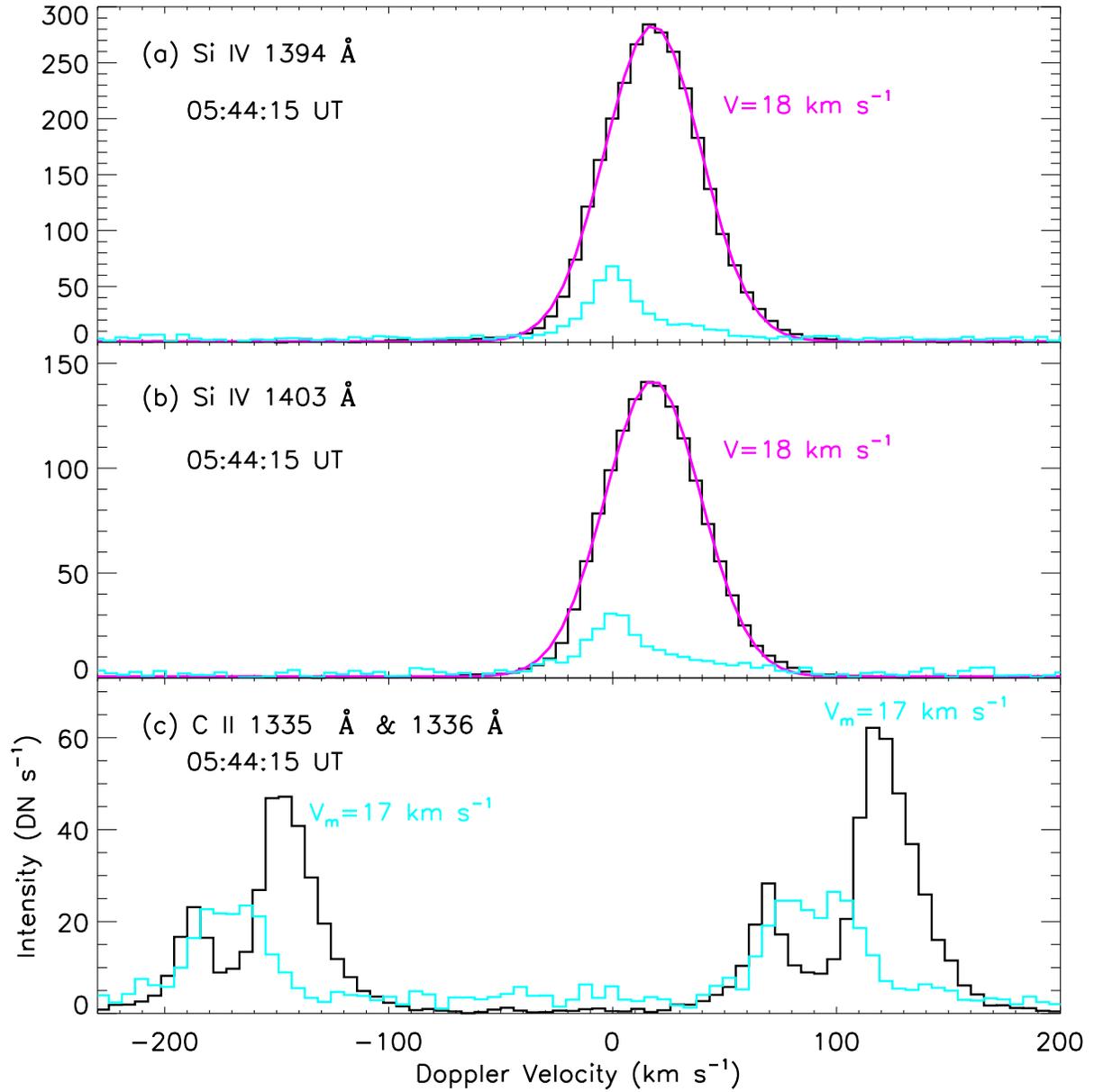}
\caption{Line spectra at the flare footpoint location (black) and
quiet region (cyan, after multiplying by 10) at $\sim$05:44:15~UT.
The magenta curves represent the single Gaussian fitting results.
The Doppler velocities of the \siiv\ and \cii\ lines are also given.
\label{line}}
\end{figure}

\begin{figure}
\centering
\includegraphics[width=\linewidth,clip=]{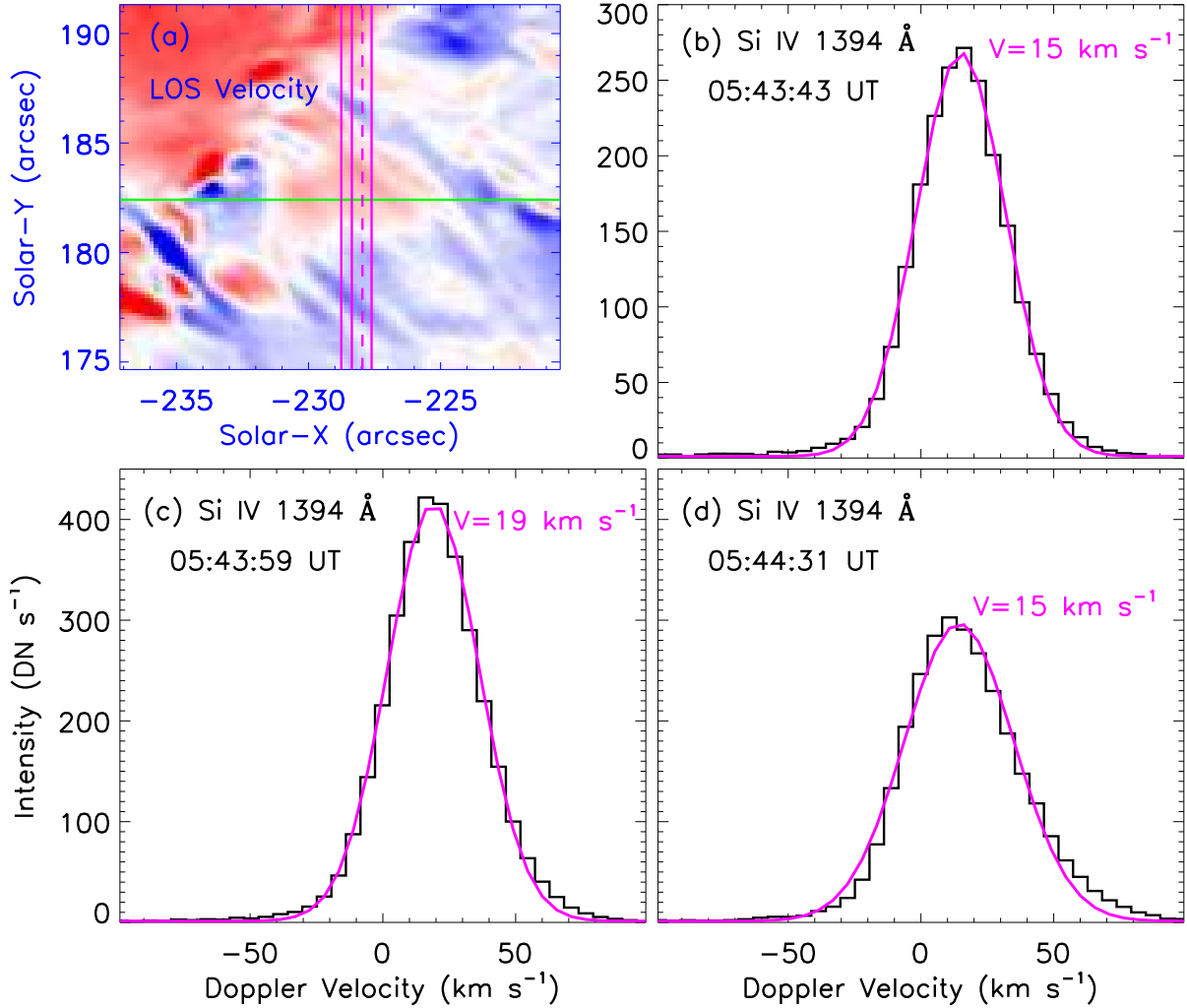}
\caption{Panel~(a): Same as Figure~\ref{spect}~(d) but with a small
FOV of $\sim$16.5\arcsec$\times$16.5\arcsec\ observed by the NVST.
The vertical magenta lines mark IRIS slits. Panels(b)$-$(d): Line
spectra for \siiv~1394~{\AA} (black) and their single Gaussian
fitting results (magenta) at three different slits, as marked by the
vertical solid lines in panel~(a). The Doppler velocities are also
given. \label{line1}}
\end{figure}

To look closely the Doppler velocity, we then extract the spectral
line profiles from the IRIS spectra at the footpoint (Ft2) location
nearby the flare ribbon, as shown in Figure~\ref{line}. Panels~(a)
\& (b) plot the line spectral profiles at IRIS FUV windows of
`\siiv~1394~{\AA}' and `\siiv~1403~{\AA}', containing the transition
region lines of \siiv, which are widely applied to study the
chromospheric condensation
\citep[e.g.,][]{Battaglia15,Brannon16,Brosius17,Li17,Tian18,Yu20}.
Similar to previous findings \citep{Zhang16b,Li18,Li19}, each line
spectrum is isolated, showing a good Gaussian profile at the site of
the flare footpoint. Therefore, each observed line profile of \siiv\
is fitted by a single Gaussian function superimposed on a constant
background. The fitting result for each line profile is overplotted
with a magenta curve, which corresponds well with the raw line
profile. Panel~(c) presents the line spectral profile at
`\cii~1336~{\AA}' window, which contains the double \cii\ lines
formed in the lower transition region. It can be seen that they appear
central-reversed line profiles. Thus, the moment analysis
\citep[e.g.,][]{Li19,Yu20} rather than the Gaussian fitting is
applied to the observed \cii\ line profile. The Doppler velocity of
each spectral line is determined by its real line center subtracting
the rest wavelength. The real line centers of \siiv\ and \cii\ lines
are identified from the single Gaussian fitting and moment analysis
method, respectively. The rest wavelengths are determined from their
line profiles in the quiet region. To improve the signal-to-noise
ratio, the line profile in the quiet region is averaged over 10
pixels between about Y$\approx$172.1\arcsec\ and
Y$\approx$175.4\arcsec\ along the IRIS slit, as outlined by the two
dashed green lines in Figure~\ref{spect}. It should be pointed out
that the line profiles in the quiet region are multiplied by 10, so
that they can be clearly seen in the same window with the flare
footpoint line spectra. Their Doppler velocities are easy to be
estimated from their line centers and rest wavelengths, which are
roughly equal to 18~km~s$^{-1}$ for the \siiv\ lines and
17~km~s$^{-1}$ for the \cii\ lines, as given in Figure~\ref{line}.
In order to provide the spectral information at the
different footpoint locations, we plot three other IRIS line spectra
at the window of `\siiv~1394~{\AA}' before and after 05:44:15~UT, as
shown in Figure~\ref{line1}~(b)$-$(d). Their slit positions are
marked by the solid vertical lines in panel~(a), which shows the
zoomed LOS velocity image in the H$\alpha$ waveband. Similar to
previous results, each observed line profiles (black) is isolated,
revealing a good Gaussian profile. Therefore, it is fitted with a
single Gaussian function superimposed on a constant background, as
indicated by the overplotted magenta curve in each panel. The
Doppler velocities are estimated to be about 15$-$19~km~s$^{-1}$,
and they all show red-shifted velocities.

The red shifts of cool lines at the flare footpoint locations could
be regarded as downflows caused by the chromospheric condensation
\citep[e.g.,][]{Zhang16a,Libbrecht19,Graham20,Yu20}. However, we
could not find blue shifts of the hot line (i.e., \fexxi) at the
same footpoint location, largely due to the IRIS raster mode, for
instance, the IRIS slit only scanned the edge of the footpoint
location, where it was very weak and could not generate the hot
(e.g., 10~MK) flare line. On the other hand, the mass converging
motions driven by the chromospheric evaporation are also expected to
be seen from double footpoints to the loop top along the new formed
flare loop. Figure~\ref{image} shows that a hot flare loop in XRT
Al\_mesh, AIA~131~{\AA}, and 94~{\AA} connects double footpoints,
which are overlayed on two ribbons in SJI~1400~{\AA}, H$\alpha$ line
core, and AIA~1600~{\AA}. The double footpoints rooted in the
positive (Ft1) and negative (Ft2) polarities, respectively. They
approach each other along the flare loop, merging together into one
single source, as shown in Figure~\ref{image}~(h). To study the mass
motions in detail, we plot the time-distance diagrams along the hot
flare loop in EUV images at a high temperature, such as
AIA~131~{\AA}, 94~{\AA}, and 335~{\AA}. Figure~\ref{slice} presents
the time-distance diagrams from 05:42:30~UT to 05:48:30~UT in these
three wavebands. To clearly display the mass converging motions, the
running difference rather than the original data series are used
here. The converging mass motions can be obviously seen at
wavelengths of AIA~131~{\AA}, 94~{\AA}, and 335~{\AA}. They both
move from double footpoints (Ft1 \& Ft2) along the flare loop,
merging together at the loop top (Lt), as indicated by the magenta
arrows. During this process, the flare loop is filled up with hot
plasmas. The converging motions are strongly indicative of upflows
driven by the chromospheric evaporation. Interestingly, the onset
time of the merging process at the loop top becomes later and later
with the plasma temperature decreases, as indicated by the vertical
dashed lines. For instance, the merging process starts at about
05:44:06~UT in AIA~131~{\AA} ($\sim$10~MK), and it becomes at around
05:44:30~UT in AIA~94~{\AA} ($\sim$6~MK), while it begins at about
05:44:38~UT in AIA~335~{\AA} ($\sim$3~MK). This is strongly related
to the evaporation speeds of the converging flows, which can be
estimated by the linear fitting. Then, the 3-sigma (3$\sigma$)
uncertainty of the fitted parameter could be regarded as the error
of the derived velocity. For example, the average evaporation speeds
within error bars are estimated to be about 206$\pm$15~km~s$^{-1}$
and 210$\pm$16~km~s$^{-1}$ in AIA~131~{\AA}. They decrease to
roughly 153$\pm$9~km~s$^{-1}$ and 157$\pm$11~km~s$^{-1}$ in
AIA~94~{\AA}, and further decrease to around 126$\pm$5~km~s$^{-1}$
and 128$\pm$6~km~s$^{-1}$ in AIA~335~{\AA}. This observational fact
is consistent with previous findings, for instance, the
temperature-dependent upflows for the coronal lines during the flare
impulsive phase \citep[e.g.,][]{Milligan09,Chen10}. Meanwhile, the
converging mass motions takes place from $\sim$05:43:30~UT to
$\sim$05:44:38~UT, which is after the C6.7 flare onset time and
roughly equal to the lifetime (such as 1~minute) of the first HXR
pulse recorded by the Fermi/GBM, as shown by the overplotted light
curves. This implies that the upflows are strongly depending on the
nonthermal energy.

\begin{figure}
\centering
\includegraphics[width=\linewidth,clip=]{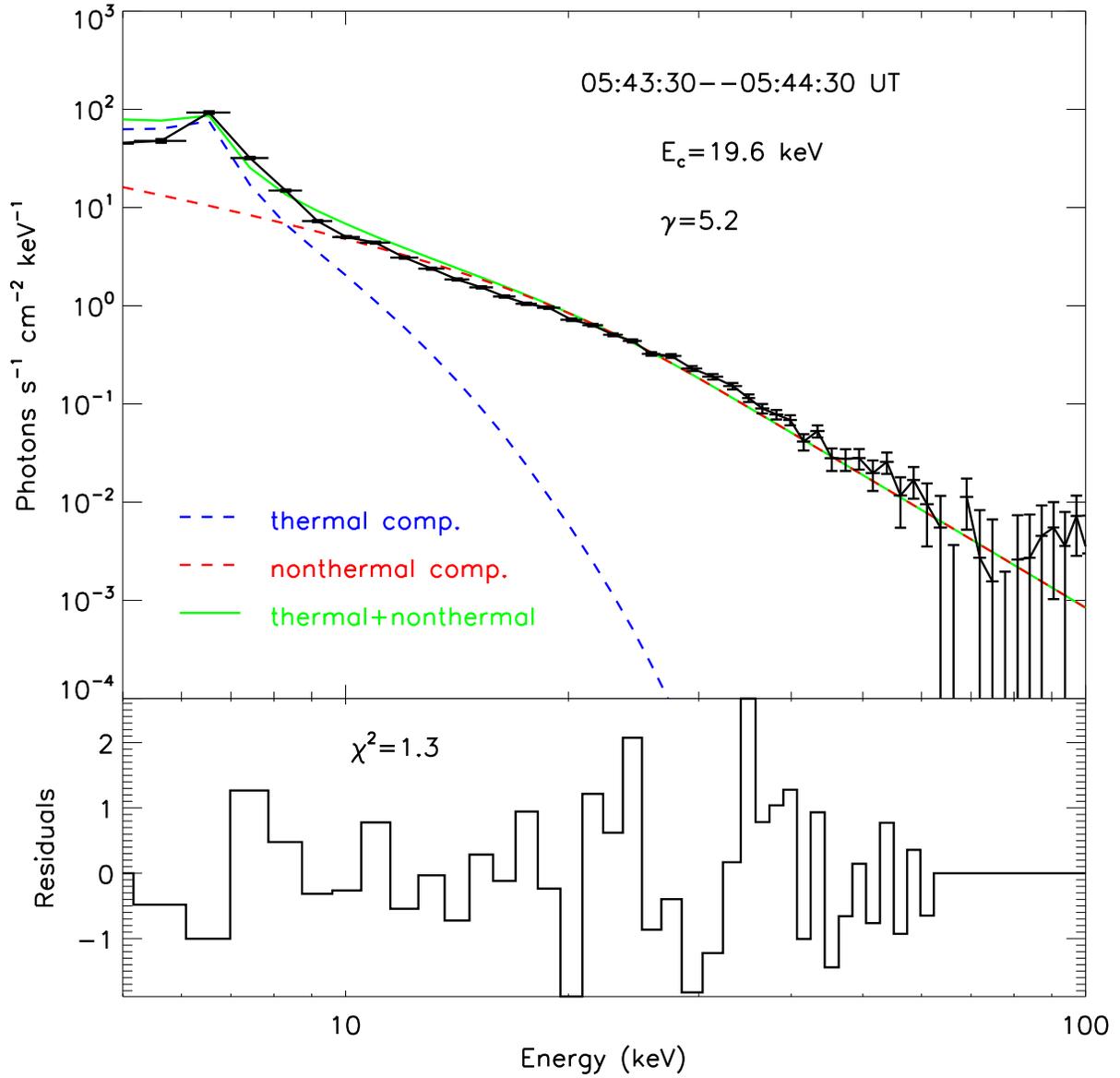}
\caption{Fermi/GBM X-ray spectrum with an integrated time of 1
minute during 05:43:30$-$05:44:30~UT, the fitted thermal (blue) and
nonthermal (red) components, as well as the sum of both components
(green). The low-energy cutoff ($E_c$), spectral index ($\gamma$),
and the Chi-squared residual ($\chi^2$) are labeled. \label{hxr}}
\end{figure}

The X-ray energy spectrum measured by the Fermi/GBM is used to
analyze the deposited energy flux during the chromospheric
evaporation, as shown in Figure~\ref{hxr}. In this study, the
two-components model such as a single thermal (vth) plus a
non-thermal thick target (thick2) is used to fit the observed
X-ray spectrum during the whole lifetime of the first HXR pulse,
i.e., from 05:43:30~UT to 05:44:30~UT. Then, the cutoff energy
($E_c$) and the spectral index ($\gamma$) of the photon spectrum can
be obtained from the fitting result. The Chi-squared residual
($\chi^2$) shows a reasonable fitting result, i.e., $\chi^{2}<3$
\citep[e.g.,][]{Sadykov15}. Thus, the nonthermal power ($P$) above
the cutoff energy of the accelerated electrons could be estimated by
using the Equation~\ref{ptot} \citep[see,][]{Aschwanden05,Zhang16b}.
\begin{equation}
 P (E \geq E_c) = 1.16 \times 10^{24} \gamma^{3} I_1 (\frac{E_c}{E_1})^{-(\gamma -1)},
 \label{ptot}
\end{equation}
where I$_1$ represents the photon count rates at energies of
E~$\geq$~E$_c$, and E$_1$ denotes the lower cutoff energy. Assuming
that $I_1~\approx$~10$^{3}$~photon~s$^{-1}$~cm$^{-2}$ for the C6.7
flare \citep{Zhang16b} and $E_c~=E_1$ \citep{Aschwanden05}, $P (E
\geq E_c)$ could be estimated to about
1.6$\times$10$^{29}$~erg~s$^{-1}$. The X-ray source area ($A$) is
estimated from the Hinode/XRT image, as outlined by the brown
contour in Figure~\ref{flux}. Considering the projection effect, it
is about 2.8$\times$10$^{18}$~cm$^{-2}$. So, the nonthermal energy
flux ($P/A$) is roughly equal to
$\sim$5.7$\times$10$^{10}$~erg~s$^{-1}$~cm$^{-2}$, which is large
enough (i.e., $>$10$^{10}$~erg~s$^{-1}$~cm$^{-2}$) to drive the
explosive evaporation \citep{Fisher85,Zarro88,Kleint16}.

\section{Conclusion and Discussion}
\begin{figure}
\centering
\includegraphics[width=\linewidth,clip=]{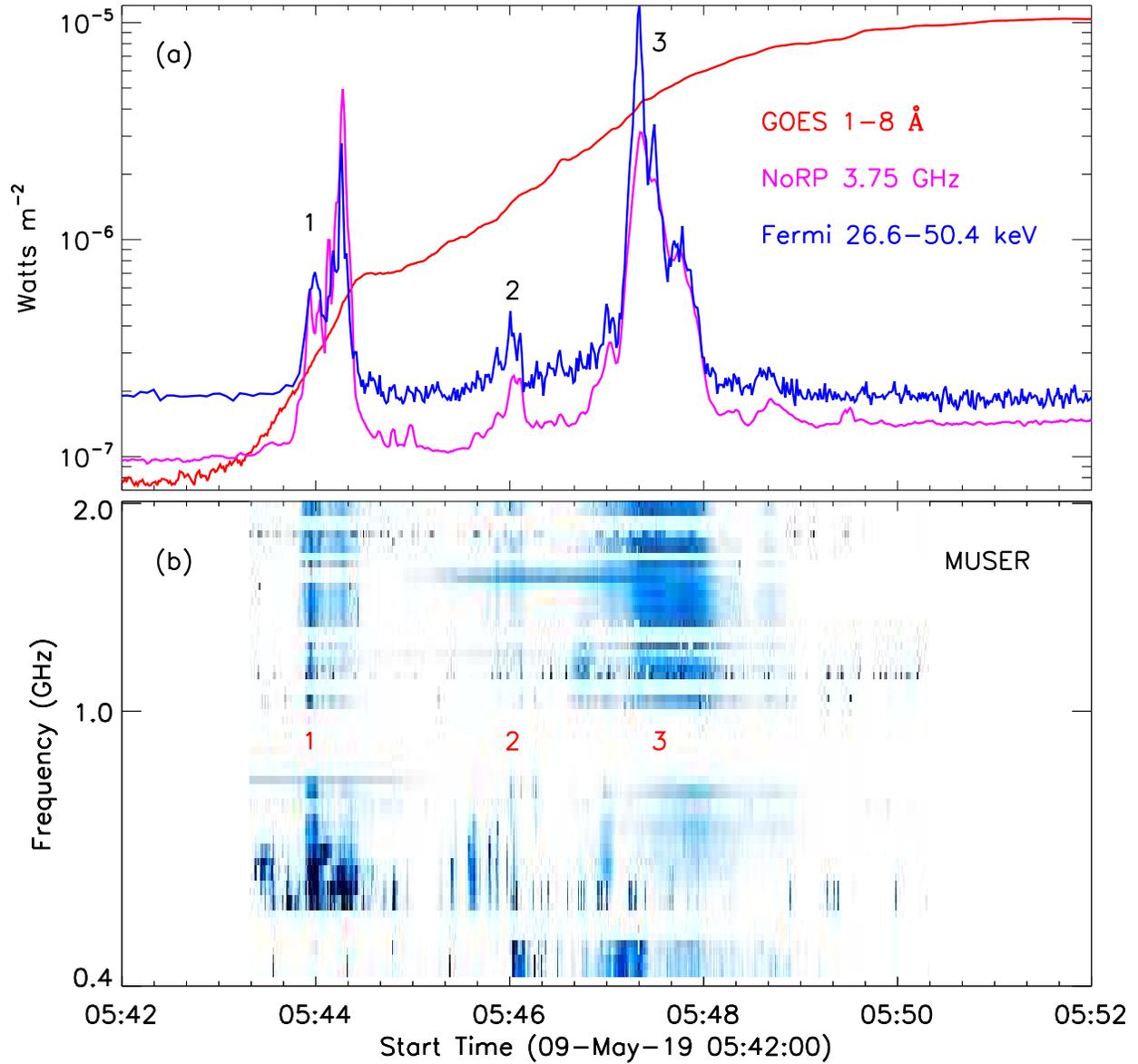}
\caption{Overview of the solar flare on 9 May 2019. Panel~(a):
Full-disk light curves from 05:42~UT to 05:52~UT, recorded by the
GOES~1$-$8~{\AA} (red), NoRP~3.75~GHz (magenta), and
Fermi~26.6$-$50.4~keV (blue), respectively. Panel~(b): The radio
dynamic spectrum observed by the MUSER. The Roman numerals mark
three nonthermal pulses. \label{flux}}
\end{figure}

\begin{figure}
\centering
\includegraphics[width=\linewidth,clip=]{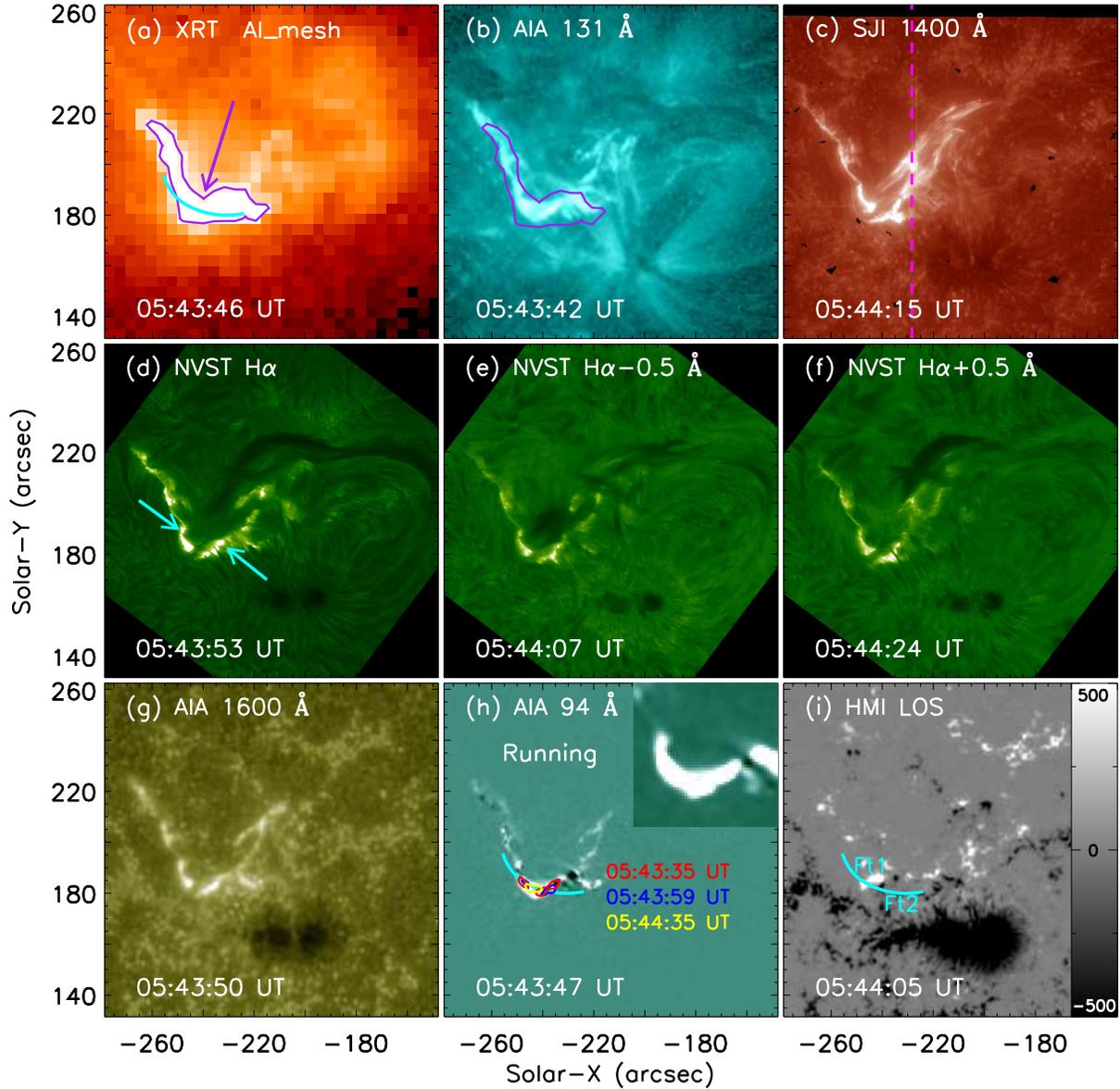}
\caption{Multi-wavelength snapshots with a FOV of about
130\arcsec$\times$130\arcsec\ of the C6.7 flare observed by the
Hinode/XRT (a), IRIS/SJI (c), the NVST (d$-$f), SDO/AIA (b, g, h),
and SDO/HMI (i). The purple profile is made from the XRT image, and
the contour level is set at 80\% of the maximum intensity. The
purple arrow indicates the flare loop, and two cyan arrows mark
flare ribbons. The vertical magenta line represents the IRIS slit.
Panel~(h) presents the running difference map, and the overplotted
contours are EUV emissions in AIA~94~{\AA} at the level of 60\% at
three fixed instances of time. The zoomed image shows the
AIA~94~{\AA} running difference map with a small FOV of about
30\arcsec$\times$30\arcsec. The cyan curve marks the flare loop used
to perform the time-distance diagrams in Figure~\ref{slice}.
\label{image}}
\end{figure}

\begin{figure}
\centering
\includegraphics[width=\linewidth,clip=]{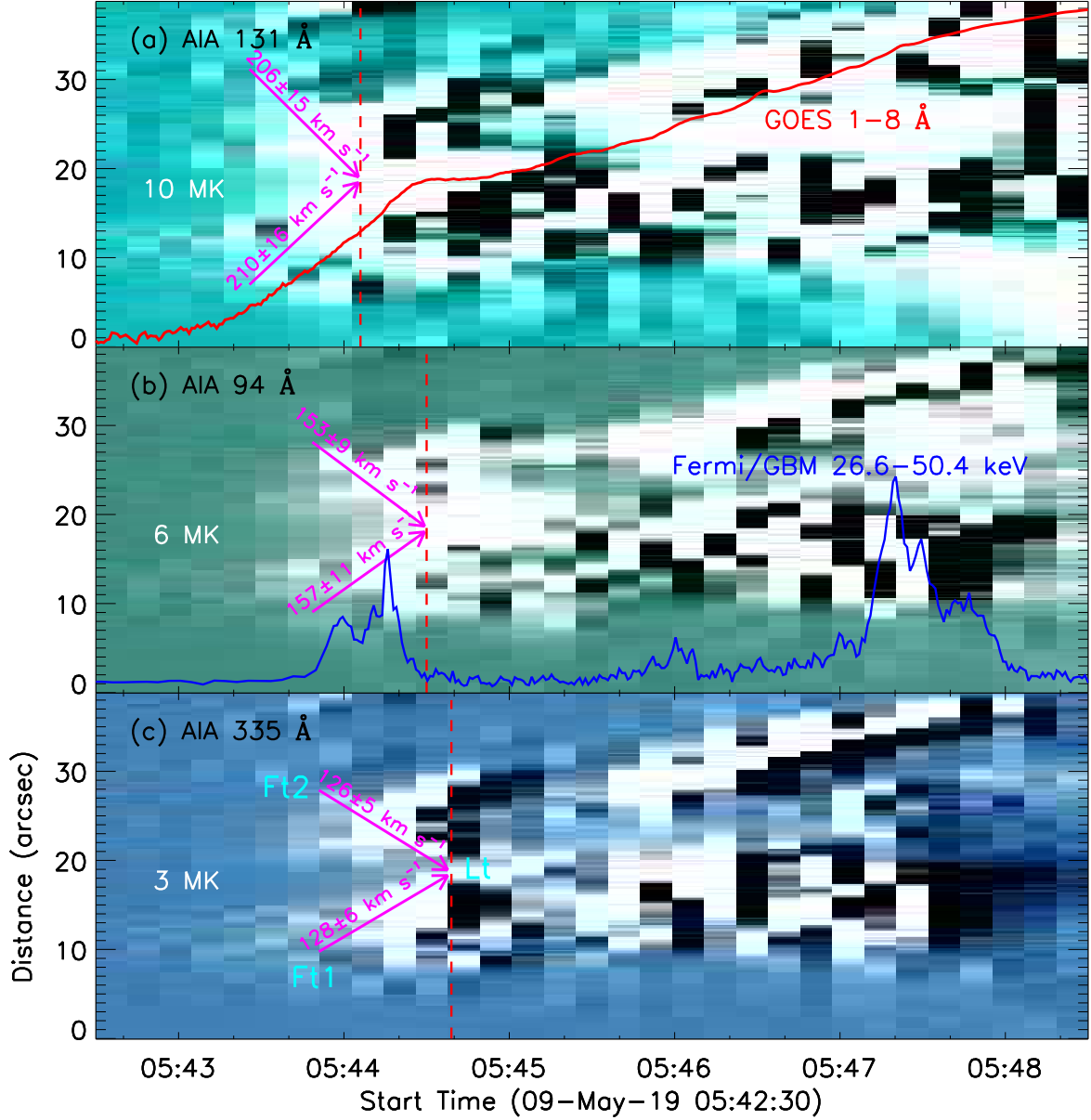}
\caption{Time-distance diagrams along the flare loop, made from the
AIA data series in 131~{\AA} (a), 94~{\AA} (b), and 335~{\AA} (c),
respectively. The magenta arrows indicate the upflows from double
footpoints (Ft), and the vertical dashed lines mark the onset time
of merging processes at the loop-top (Lt) region. The overplotted
line curves are the GOES~1$-$8~{\AA} (red) and Fermi~26.6$-$50.4~keV
(blue) fluxes. \label{slice}}
\end{figure}

Using the multi-wavelength observations measured by the NVST, IRIS,
SDO/AIA, Fermi/GBM, Hinode/XRT, MUSER, and NoRP, we explored
simultaneous observations of chromospheric evaporation and
condensation in a C6.7 flare on 9 May 2019. The cool lines such as
\siiv\ and \cii\ measured by the IRIS show obviously red shifted
velocities at slow speeds of about 15$-$19~km~s$^{-1}$ at one flare
footpoint. The same red shifts can be seen in the LOS velocity image
at double footpoints, which is derived from the two H$\alpha$ off
bands ($\pm$0.5~{\AA}) observed by the NVST. Those red
shifts at slow speeds in the solar chromosphere and transition
region indicate the low-velocity downflows driven by the
chromospheric condensation
\citep[e.g.,][]{Zhang16a,Libbrecht19,Yu20}. The high-temperature
X-ray/EUV images detected by the Hinode/XRT and SDO/AIA show a hot
flare loop, within double footpoints rooting in the positive and
negative magnetic fields, respectively. The converging mass motions
can be clearly seen from double footpoints to the loop top at fast
speeds of 126$-$210~km~s$^{-1}$, which could be the indicative of
high-velocity upflows driven by the chromospheric evaporation
\citep{Li17,Zhang19}. It should be pointed out that the
estimated speeds from the time-distance diagram are horizontal
components. On the other hand, the high-temperature EUV emissions in
AIA~94~{\AA} appear to rise from double footpoints to the loop top
and eventually merge into a single source (see,
Figure~\ref{image}~h), which are quite similar to the converging
motion of HXR sources \citep[e.g.,][]{Liu06,Liu08,Ning09,Ning11}.
Therefore, the motion velocities derived from the time-distance
diagrams here are most possible to be the projected velocities of
the upflows driven by the chromospheric evaporation rather than the
motion of the flare loops. We would like to state that the C6.7
flare occurs closely to the solar disk center and the projection
effect is quite weak. The low-velocity downflows of cool lines and
the high-velocity upflows of hot plasmas are simultaneously observed
during the impulsive phase of the C6.7 flare, suggesting an
explosive chromospheric evaporation
\citep[e.g.,][]{Milligan06a,Tian15,Zhang16b,Brosius17,Sadykov19}.
Moreover, the nonthermal energy flux is estimated to be
$\sim$5.7$\times$10$^{10}$~erg~s$^{-1}$~cm$^{-2}$, implying a strong
heating energy of the electron beams during the flare impulsive
phase. Obviously, it is larger than the critical threshold between
the gentle and explosive evaporations presented by \cite{Fisher85}.
Meanwhile, the characteristic timescale of the chromospheric
evaporation can be estimated from the converging mass motions of hot
plasmas in Figure~\ref{slice}. It could also be estimated from the
lifetime of the first HXR/microwave pulse in Figure~\ref{flux}. They
both show a characteristic timescale of about 1 minute, which is
equal to our previous finding \citep{Li18}, and it is similar to the
typical timescale of an explosive evaporation
\citep{Zarro88,Liu08,Sadykov15}. Our observational facts demonstrate
an explosive chromospheric evaporation during the flare impulsive
phase.

It is necessary to state that the physical origin of the explosive
chromospheric evaporation. The IRIS observed the C6.7 flare in the
raster mode rather than the fixed slit, and the time cadence of NVST
is as low as about 43~s, while the HXR/micowave pulses only remains
about 1~minute. We cannot obtain the point-to-point correspondence
between the time series of Doppler velocities and HXR/micowave
fluxes, as done in previous studies \citep{Li15,Tian15,Li18}. We
cannot find the spatial correlation between the upflows and
HXR/micowave sources \citep[see,][]{Milligan09,Brosius16,Zhang16b},
because the lack of direct observation of HXR/micowave images.
However, the explosive chromospheric evaporation occurred during the
impulsive phase of the C6.7 flare, and the high-velocity upflows
show a good temporal correlation with the evolution of the first HXR
pulse observed by the Fermi/GBM, i.e, between $\sim$05:43:30~UT to
$\sim$05:44:30~UT. The first HXR pulse corresponds well to the
microwave pulse observed by the NoRP, which is also accompanied by a
type III radio burst detected by the MUSER, as shown in
Figure~\ref{flux}. Moreover, the heating energy released by electron
beams is larger than the critical threshold
\citep{Fisher85,Zarro88,Sadykov19}. Therefore, the explosive
chromospheric evaporation is most likely to be driven by nonthermal
electrons accelerated by the magnetic reconnection during the flare
impulsive phase
\citep[e.g.,][]{Tian15,Warren16,Zhang16b,Li17,Lee17}.

In this study, high-velocity upflows are identified as the fast
evaporation speeds of the mass motions in AIA high-temperature EUV
images. The evaporation speeds appear to be quite sensitive to the
plasma temperature in the AIA EUV images. The AIA~131~{\AA} emission
during the solar flare is mainly emitted from the \fexxi\ line,
which has a formation temperature of about 10~MK \citep{Lemen12},
and their evaporation speeds are estimated to be about
206$-$210~km~s$^{-1}$. The AIA~94~{\AA} radiation is from the
\fexviii\ line with a formation temperature of roughly 6~MK
\citep{Lemen12}, and their evaporation speeds are about
153$-$157~km~s$^{-1}$. The AIA~335~{\AA} radiation is from the
\fexvi\ line with a formation temperature of roughly 3~MK
\citep{Lemen12}, and the evaporation speeds are around
126$-$128~km~s$^{-1}$. Therefore, the hot plasmas originated from
double footpoints in AIA~131~{\AA} are first converging and merging
at the loop-top region, while those in AIA~335~{\AA} are last
converging and merging at the loop-top region in our observation, as
indicated by the vertical dashed lines in Figure~\ref{slice}. Our
observations are consistent with pervious findings using the
spectroscopic observations \citep{Milligan09,Chen10}. They found
that the plasma velocities at flare footpoints strongly depend on
the formation temperature of emission lines. However, we did not
find the similar evaporation motions in the lower temperature EUV
wavelengths, such as AIA~211~{\AA}, 193~{\AA}, and 171~{\AA}. This
is probably because that these AIA images are much easier to be
saturated during the solar flare. Another possibility could be due
to their lower response temperature. The evaporation motion may only
appear at the higher temperature wavelengths, i.e., $>$2~MK
\citep[see,][]{Milligan09}. It should be pointed out that there are
two peaks ($\sim$10~MK and $\sim$0.4~MK) in the profile of 131~{\AA}
temperature response function. Here, we chose the higher formation
temperature ($\sim$10~MK) as the AIA~131~{\AA} response temperature,
mainly because that the lower temperature EUV wavelengths
(171~{\AA}, 193~{\AA}, and 211~{\AA}) had very weak or even not
response for the chromospheric evaporation in the C6.7 flare. In
this study, we only investigate the explosive chromospheric
evaporation during the first HXR/microwave pulse between roughly
05:43~UT and 05:45~UT. We also notice that the HXR/microwave fluxes
show another two nonthermal pulses from about 05:45~UT to 05:49~UT,
but we do not study them. This is because that the IRIS slit leaves
the flare ribbon/footpoint, and the AIA EUV images are saturated
when the flare reaches closely to its maximum.

We wanted to stress that the HXR source area is replaced by the
X-ray source area ($A$) measured by the Hinode/XRT. This is due to
the observational limitation, since the Fermi/GBM only provided the
HXR flux during solar flares, we cannot find the HXR image for this
flare. The X-ray source measured by the Hinode/XRT contains the hot
loop seen in the AIA~131~{\AA}, and it also includes two ribbons
observed in the NVST H$\alpha$ image, as shown in
Figure~\ref{image}. It should be larger than the HXR source.
Therefore, the X-ray source area could be regarded as the upper
limit of the HXR area, and it is indeed a bit larger than the
previous finding for a C6.5-class flare \citep[e,g.,][]{Zhang16b}.
However, this does not affect our results, since the nonthermal
energy flux is based on the minimum estimation, but it is still
larger than the typical threshold of the explosive evaporation. On
the other hand, it is well known that the spectral lines for \siiv\
and \cii\ are optically thin and thick, respectively. Therefore,
their Doppler velocities are determined by the different methods,
for instance, the single Gaussian fitting for the \siiv\ lines,
while the moment analysis technique for the \cii\ lines
\citep{Li19,Yu20}.

\acknowledgments We would like to thank the anonymous
referee for helping to improve the quality of the paper. We thank
the teams of NVST, IRIS, SDO, MUSER, Fermi, Hinode/XRT, NoRP, and
GOES for their open data use policy. This study is supported by NSFC
under grant 11973092, 12073081, 11873095, 11790302, the Strategic
Priority Research Program on Space Science, CAS, Grant No.
XDA15052200 and XDA15320301. D.~L. is also supported by the Surface
Project of Jiangsu Province (BK20211402). The Laboratory No.
2010DP173032.

\begin{table}
\caption{Details of observational instruments used in this paper.}
\centering
\begin{tabular}{c c c c c c c}
\hline
Instruments &  Wavebands          &  Cadence      &     Pixel scale        &   Descriptions        \\
\hline
            &  1394 \& 1403~{\AA} &  $\sim$16.2~s &  $\sim$0.33\arcsec     &   \siiv\ lines        \\
IRIS        &  1335 \& 1336~{\AA} &  $\sim$16.2~s &  $\sim$0.33\arcsec     &   \cii\  lines        \\
            &  SJI~1400~{\AA}     &  $\sim$65~s  &  $\sim$0.33\arcsec      &    FUV image          \\
\hline
            &  6562.8~{\AA}       &  $\sim$43~s   &   $\sim$0.165\arcsec   &   H$\alpha$ line core  \\
NVST        &  $\pm$0.5~{\AA}     &  $\sim$43~s   &   $\sim$0.165\arcsec   &   H$\alpha$ off bands  \\
\hline
            & AIA~94, 131 \& 335~{\AA} &    12~s  &    0.6\arcsec          &   EUV image            \\
SDO         &  AIA~1600~{\AA}     &    24~s       &    0.6\arcsec          &   UV image             \\
            &  HMI LOS            &    45~s       &    0.6\arcsec          &   Magnetogram          \\
\hline
Hinode/XRT  & Al\_mesh            &     --        &  $\sim$4.1\arcsec      &    X-ray image         \\
\hline
GOES-16     &  1$-$8~{\AA}        &    1~s        &         -             &    SXR flux            \\
\hline
Fermi/GBM   & 26.6$-$50.4~keV     & $\sim$4.096~s &         -             &    HXR flux            \\
            & 5$-$100~keV         & $\sim$4.096~s &         -             &    X-ray spectrum      \\
\hline
MUSER       &  0.4$-$2~GHz        &     0.025~s   &         -             &   Radio dynamic spectrum \\
\hline
NoRP        &  3.75~GHz           &      1~s      &         -             &   Microwave flux   \\
\hline
\end{tabular}
\label{tab1}
\end{table}

\begin{thebibliography}{}
\bibitem[Asai et al.(2006)]{Asai06} Asai, A., Nakajima, H., Shimojo, M., et al.\ 2006, \pasj, 58, L1. doi:10.1093/pasj/58.1.L1
\bibitem[Aschwanden \& Benz(1995)]{Aschwanden95} Aschwanden, M.~J. \& Benz, A.~O.\ 1995, \apj, 438, 997. doi:10.1086/175141
\bibitem[Aschwanden(2005)]{Aschwanden05} Aschwanden, M.~J.\ 2005, Physics of the Solar Corona. An Introduction with Problems and Solutions (2nd edition), by M.J. Aschwanden.  892 pages.  ISBN 3-540-30765-6, Library of Congress Control Number: 2005937065.  Praxis Publishing Ltd., Chichester, UK; Springer, New York, Berlin, 2005.
\bibitem[Ashfield \& Longcope(2021)]{Ashfield21} Ashfield, W.~H. \& Longcope, D.~W.\ 2021, \apj, 912, 25. doi:10.3847/1538-4357/abedb4
\bibitem[Battaglia et al.(2015)]{Battaglia15} Battaglia, M., Kleint, L., Krucker, S., et al.\ 2015, \apj, 813, 113. doi:10.1088/0004-637X/813/2/113
\bibitem[Benz(2017)]{Benz17} Benz, A.~O.\ 2017, Living Reviews in Solar Physics, 14, 2. doi:10.1007/s41116-016-0004-3
\bibitem[Brannon(2016)]{Brannon16} Brannon, S.~R.\ 2016, \apj, 833, 101. doi:10.3847/1538-4357/833/1/101
\bibitem[Brosius \& Holman(2010)]{Brosius10} Brosius, J.~W. \& Holman, G.~D.\ 2010, \apj, 720, 1472. doi:10.1088/0004-637X/720/2/1472
\bibitem[Brosius(2013)]{Brosius13} Brosius, J.~W.\ 2013, \apj, 762, 133. doi:10.1088/0004-637X/762/2/133
\bibitem[Brosius \& Daw(2015)]{Brosius15} Brosius, J.~W. \& Daw, A.~N.\ 2015, \apj, 810, 45. doi:10.1088/0004-637X/810/1/45
\bibitem[Brosius et al.(2016)]{Brosius16} Brosius, J.~W., Daw, A.~N., \& Inglis, A.~R.\ 2016, \apj, 830, 101. doi:10.3847/0004-637X/830/2/101
\bibitem[Brosius \& Inglis(2017)]{Brosius17} Brosius, J.~W. \& Inglis, A.~R.\ 2017, \apj, 848, 39. doi:10.3847/1538-4357/aa8a68
\bibitem[Brosius \& Inglis(2018)]{Brosius18} Brosius, J.~W. \& Inglis, A.~R.\ 2018, \apj, 867, 85. doi:10.3847/1538-4357/aae5f5
\bibitem[Carmichael(1964)]{Carmichael64} Carmichael, H.\ 1964, NASA Special Publication, 451
\bibitem[Chen \& Ding(2010)]{Chen10} Chen, F. \& Ding, M.~D.\ 2010, \apj, 724, 640. doi:10.1088/0004-637X/724/1/640
\bibitem[Chen et al.(2017)]{Chen17} Chen, Y., Wu, Z., Liu, W., et al.\ 2017, \apj, 843, 8. doi:10.3847/1538-4357/aa7462
\bibitem[Chen et al.(2020)]{Chen20} Chen, B., Shen, C., Gary, D.~E., et al.\ 2020, Nature Astronomy, 4, 1140. doi:10.1038/s41550-020-1147-7
\bibitem[Czaykowska et al.(1999)]{Czaykowska99} Czaykowska, A., De Pontieu, B., Alexander, D., et al.\ 1999, \apjl, 521, L75. doi:10.1086/312176
\bibitem[Czaykowska et al.(2001)]{Czaykowska01} Czaykowska, A., Alexander, D., \& De Pontieu, B.\ 2001, \apj, 552, 849. doi:10.1086/320553
\bibitem[De Pontieu et al.(2014)]{Dep14} De Pontieu, B., Title, A.~M., Lemen, J.~R., et al.\ 2014, \solphys, 289, 2733. doi:10.1007/s11207-014-0485-y
\bibitem[De Pontieu et al.(2021)]{Dep21} De Pontieu, B., Polito, V., Hansteen, V., et al.\ 2021, \solphys, 296, 84. doi:10.1007/s11207-021-01826-0
\bibitem[Ding et al.(1996)]{Ding96} Ding, M.~D., Watanabe, T., Shibata, K., et al.\ 1996, \apj, 458, 391. doi:10.1086/176822
\bibitem[Dominique et al.(2018)]{Dominique18} Dominique, M., Zhukov, A.~N., Heinzel, P., et al.\ 2018, \apjl, 867, L24. doi:10.3847/2041-8213/aaeace
\bibitem[Doschek et al.(2013)]{Doschek13} Doschek, G.~A., Warren, H.~P., \& Young, P.~R.\ 2013, \apj, 767, 55. doi:10.1088/0004-637X/767/1/55
\bibitem[Dud{\'\i}k et al.(2016)]{Dudik16} Dud{\'\i}k, J., Polito, V., Janvier, M., et al.\ 2016, \apj, 823, 41. doi:10.3847/0004-637X/823/1/41
\bibitem[Falewicz et al.(2009)]{Falewicz09} Falewicz, R., Rudawy, P., \& Siarkowski, M.\ 2009, \aap, 508, 971. doi:10.1051/0004-6361/200912781
\bibitem[Fletcher \& Hudson(2008)]{Fletcher08} Fletcher, L. \& Hudson, H.~S.\ 2008, \apj, 675, 1645. doi:10.1086/527044
\bibitem[Fletcher et al.(2011)]{Fletcher11} Fletcher, L., Dennis, B.~R., Hudson, H.~S., et al.\ 2011, \ssr, 159, 19. doi:10.1007/s11214-010-9701-8
\bibitem[Fisher et al.(1985)]{Fisher85} Fisher, G.~H., Canfield, R.~C., \& McClymont, A.~N.\ 1985, \apj, 289, 414. doi:10.1086/162901
\bibitem[Golub et al.(2007)]{Golub07} Golub, L., Deluca, E., Austin, G., et al.\ 2007, \solphys, 243, 63. doi:10.1007/s11207-007-0182-1
\bibitem[Graham \& Cauzzi(2015)]{Graham15} Graham, D.~R. \& Cauzzi, G.\ 2015, \apjl, 807, L22. doi:10.1088/2041-8205/807/2/L22
\bibitem[Graham et al.(2020)]{Graham20} Graham, D.~R., Cauzzi, G., Zangrilli, L., et al.\ 2020, \apj, 895, 6. doi:10.3847/1538-4357/ab88ad
\bibitem[Hirayama(1974)]{Hirayama74} Hirayama, T.\ 1974, \solphys, 34, 323. doi:10.1007/BF00153671
\bibitem[Hong et al.(2021)]{Hong21} Hong, Z., Li, D., Zhang, M., et al.\ 2021, \solphys, 296, 171. doi:10.1007/s11207-021-01922-1
\bibitem[Innes et al.(1997)]{Innes97} Innes, D.~E., Inhester, B., Axford, W.~I., et al.\ 1997, \nat, 386, 811. doi:10.1038/386811a0
\bibitem[Jiang et al.(2021)]{Jiang21} Jiang, C., Feng, X., Liu, R., et al.\ 2021, arXiv:2107.08204
\bibitem[Kamio et al.(2005)]{Kamio05} Kamio, S., Kurokawa, H., Brooks, D.~H., et al.\ 2005, \apj, 625, 1027. doi:10.1086/429749
\bibitem[Karlicky(1998)]{Karlicky98} Karlicky, M.\ 1998, \aap, 338, 1084
\bibitem[Kleint et al.(2016)]{Kleint16} Kleint, L., Heinzel, P., Judge, P., et al.\ 2016, \apj, 816, 88. doi:10.3847/0004-637X/816/2/88
\bibitem[Kopp \& Pneuman(1976)]{Kopp76} Kopp, R.~A. \& Pneuman, G.~W.\ 1976, \solphys, 50, 85. doi:10.1007/BF00206193
\bibitem[Krucker et al.(2008)]{Krucker08} Krucker, S., Battaglia, M., Cargill, P.~J., et al.\ 2008, \aapr, 16, 155. doi:10.1007/s00159-008-0014-9
\bibitem[Lee et al.(2017)]{Lee17} Lee, K.-S., Imada, S., Watanabe, K., et al.\ 2017, \apj, 836, 150. doi:10.3847/1538-4357/aa5b8b
\bibitem[Lemen et al.(2012)]{Lemen12} Lemen, J.~R., Title, A.~M., Akin, D.~J., et al.\ 2012, \solphys, 275, 17. doi:10.1007/s11207-011-9776-8
\bibitem[Li \& Ding(2004)]{Li04} Li, J.~P. \& Ding, M.~D.\ 2004, \apj, 606, 583. doi:10.1086/382860
\bibitem[Li \& Ding(2011)]{Li11} Li, Y. \& Ding, M.~D.\ 2011, \apj, 727, 98. doi:10.1088/0004-637X/727/2/98
\bibitem[Li et al.(2012)]{Li12} Li, Y., Qiu, J., \& Ding, M.~D.\ 2012, \apj, 758, 52. doi:10.1088/0004-637X/758/1/52
\bibitem[Li et al.(2015)]{Li15} Li, D., Ning, Z.~J., \& Zhang, Q.~M.\ 2015, \apj, 813, 59. doi:10.1088/0004-637X/813/1/59
\bibitem[Li et al.(2016)]{Li16} Li, L., Zhang, J., Peter, H., et al.\ 2016, Nature Physics, 12, 847. doi:10.1038/nphys3768
\bibitem[Li et al.(2017)]{Li17} Li, D., Ning, Z.~J., Huang, Y., et al.\ 2017, \apjl, 841, L9. doi:10.3847/2041-8213/aa71b0
\bibitem[Li et al.(2018)]{Li18} Li, D., Li, Y., Su, W., et al.\ 2018, \apj, 854, 26. doi:10.3847/1538-4357/aaa9c0
\bibitem[Li et al.(2019)]{Li19} Li, Y., Ding, M.~D., Hong, J., et al.\ 2019, \apj, 879, 30. doi:10.3847/1538-4357/ab245a
\bibitem[Li(2019)]{Lid19} Li, D.\ 2019, Research in Astronomy and Astrophysics, 19, 067. doi:10.1088/1674-4527/19/5/67
\bibitem[Li et al.(2021)]{Li21} Li, D., Warmuth, A., Lu, L., et al.\ 2021, Research in Astronomy and Astrophysics, 21, 066. doi:10.1088/1674-4527/21/3/66
\bibitem[Libbrecht et al.(2019)]{Libbrecht19} Libbrecht, T., de la Cruz Rodr{\'\i}guez, J., Danilovic, S., et al.\ 2019, \aap, 621, A35. doi:10.1051/0004-6361/201833610
\bibitem[Lin et al.(2005)]{Lin05} Lin, J., Ko, Y.-K., Sui, L., et al.\ 2005, \apj, 622, 1251. doi:10.1086/428110
\bibitem[Liu et al.(2013)]{Liu13} Liu, W., Chen, Q., \& Petrosian, V.\ 2013, \apj, 767, 168. doi:10.1088/0004-637X/767/2/168
\bibitem[Liu et al.(2006)]{Liu06} Liu, W., Liu, S., Jiang, Y.~W., et al.\ 2006, \apj, 649, 1124. doi:10.1086/506268
\bibitem[Liu et al.(2008)]{Liu08} Liu, W., Petrosian, V., Dennis, B.~R., et al.\ 2008, \apj, 676, 704. doi:10.1086/527538
\bibitem[Liu et al.(2014)]{Liu14} Liu, Z., Xu, J., Gu, B.-Z., et al.\ 2014, Research in Astronomy and Astrophysics, 14, 705-718. doi:10.1088/1674-4527/14/6/009
\bibitem[Lohmann et al.(1983)]{Lohmann83} Lohmann, A.~W., Weigelt, G., \& Wirnitzer, B.\ 1983, \ao, 22, 4028. doi:10.1364/AO.22.004028
\bibitem[Loto'aniu et al.(2017)]{Lotoaniu17} Loto'aniu, P., Rodriguez, J., Redmon, R., et al.\ 2017, EGU General Assembly Conference Abstracts
\bibitem[Lysenko et al.(2020)]{Lysenko20} Lysenko, A.~L., Frederiks, D.~D., Fleishman, G.~D., et al.\ 2020, Physics Uspekhi, 63, 818. doi:10.3367/UFNe.2019.06.038757
\bibitem[Mann \& Warmuth(2011)]{Mann11} Mann, G. \& Warmuth, A.\ 2011, \aap, 528, A104. doi:10.1051/0004-6361/201014389
\bibitem[Masuda et al.(1994)]{Masuda94} Masuda, S., Kosugi, T., Hara, H., et al.\ 1994, \nat, 371, 495. doi:10.1038/371495a0
\bibitem[Meegan et al.(2009)]{Meegan09} Meegan, C., Lichti, G., Bhat, P.~N., et al.\ 2009, \apj, 702, 791. doi:10.1088/0004-637X/702/1/791
\bibitem[Milligan et al.(2006a)]{Milligan06a} Milligan, R.~O., Gallagher, P.~T., Mathioudakis, M., et al.\ 2006a, \apjl, 638, L117. doi:10.1086/500555
\bibitem[Milligan et al.(2006b)]{Milligan06b} Milligan, R.~O., Gallagher, P.~T., Mathioudakis, M., et al.\ 2006b, \apjl, 642, L169. doi:10.1086/504592
\bibitem[Milligan \& Dennis(2009)]{Milligan09} Milligan, R.~O. \& Dennis, B.~R.\ 2009, \apj, 699, 968. doi:10.1088/0004-637X/699/2/968
\bibitem[Nakajima et al.(1985)]{Nakajima85} Nakajima, H., Sekiguchi, H., Sawa, M., et al.\ 1985, \pasj, 37, 163
\bibitem[Ning et al.(2009)]{Ning09} Ning, Z., Cao, W., Huang, J., et al.\ 2009, \apj, 699, 15. doi:10.1088/0004-637X/699/1/15
\bibitem[Ning \& Cao(2010)]{Ning10} Ning, Z. \& Cao, W.\ 2010, \apj, 717, 1232. doi:10.1088/0004-637X/717/2/1232
\bibitem[Ning \& Cao(2011)]{Ning11a} Ning, Z. \& Cao, W.\ 2011, \solphys, 269, 283. doi:10.1007/s11207-010-9693-2
\bibitem[Ning(2011)]{Ning11} Ning, Z.\ 2011, \solphys, 273, 81. doi:10.1007/s11207-011-9833-3
\bibitem[Pesnell et al.(2012)]{Pesnell12} Pesnell, W.~D., Thompson, B.~J., \& Chamberlin, P.~C.\ 2012, \solphys, 275, 3. doi:10.1007/s11207-011-9841-3
\bibitem[Polito et al.(2017)]{Polito17} Polito, V., Del Zanna, G., Valori, G., et al.\ 2017, \aap, 601, A39. doi:10.1051/0004-6361/201629703
\bibitem[Polito et al.(2018)]{Polito18} Polito, V., Testa, P., Allred, J., et al.\ 2018, \apj, 856, 178. doi:10.3847/1538-4357/aab49e
\bibitem[Priest \& Forbes(2002)]{Priest02} Priest, E.~R. \& Forbes, T.~G.\ 2002, \aapr, 10, 313. doi:10.1007/s001590100013
\bibitem[Raftery et al.(2009)]{Raftery09} Raftery, C.~L., Gallagher, P.~T., Milligan, R.~O., et al.\ 2009, \aap, 494, 1127. doi:10.1051/0004-6361:200810437
\bibitem[Reep et al.(2015)]{Reep15} Reep, J.~W., Bradshaw, S.~J., \& Alexander, D.\ 2015, \apj, 808, 177. doi:10.1088/0004-637X/808/2/177
\bibitem[Reep \& Russell(2016)]{Reep16} Reep, J.~W. \& Russell, A.~J.~B.\ 2016, \apjl, 818, L20. doi:10.3847/2041-8205/818/1/L20
\bibitem[Rubio da Costa et al.(2015)]{Rubio15} Rubio da Costa, F., Liu, W., Petrosian, V., et al.\ 2015, \apj, 813, 133. doi:10.1088/0004-637X/813/2/133
\bibitem[Sadykov et al.(2015)]{Sadykov15} Sadykov, V.~M., Vargas Dominguez, S., Kosovichev, A.~G., et al.\ 2015, \apj, 805, 167. doi:10.1088/0004-637X/805/2/167
\bibitem[Sadykov et al.(2019)]{Sadykov19} Sadykov, V.~M., Kosovichev, A.~G., Sharykin, I.~N., et al.\ 2019, \apj, 871, 2. doi:10.3847/1538-4357/aaf6b0
\bibitem[Schou et al.(2012)]{Schou12} Schou, J., Scherrer, P.~H., Bush, R.~I., et al.\ 2012, \solphys, 275, 229. doi:10.1007/s11207-011-9842-2
\bibitem[Shibata \& Magara(2011)]{Shibata11} Shibata, K. \& Magara, T.\ 2011, Living Reviews in Solar Physics, 8, 6. doi:10.12942/lrsp-2011-6
\bibitem[Sturrock \& Coppi(1964)]{Sturrock64} Sturrock, P.~A. \& Coppi, B.\ 1964, \nat, 204, 61. doi:10.1038/204061a0
\bibitem[Sturrock(1966)]{Sturrock66} Sturrock, P.~A.\ 1966, \nat, 211, 695. doi:10.1038/211695a0
\bibitem[Song et al.(2018)]{Song18} Song, Y.~L., Guo, Y., Tian, H., et al.\ 2018, \apj, 854, 64. doi:10.3847/1538-4357/aaa7f1
\bibitem[Su et al.(2013)]{Su13} Su, Y., Veronig, A.~M., Holman, G.~D., et al.\ 2013, Nature Physics, 9, 489. doi:10.1038/nphys2675
\bibitem[Sui \& Holman(2003)]{Sui03} Sui, L. \& Holman, G.~D.\ 2003, \apjl, 596, L251. doi:10.1086/379343
\bibitem[Tan et al.(2020)]{Tan20} Tan, B.-L., Yan, Y., Li, T., et al.\ 2020, Research in Astronomy and Astrophysics, 20, 090. doi:10.1088/1674-4527/20/6/90
\bibitem[Temmer et al.(2007)]{Temmer07} Temmer, M., Veronig, A.~M., Vr{\v{s}}nak, B., et al.\ 2007, \apj, 654, 665. doi:10.1086/509634
\bibitem[Teriaca et al.(2003)]{Teriaca03} Teriaca, L., Falchi, A., Cauzzi, G., et al.\ 2003, \apj, 588, 596. doi:10.1086/373946
\bibitem[Tian et al.(2014)]{Tian14} Tian, H., Li, G., Reeves, K.~K., et al.\ 2014, \apjl, 797, L14. doi:10.1088/2041-8205/797/2/L14
\bibitem[Tian et al.(2015)]{Tian15} Tian, H., Young, P.~R., Reeves, K.~K., et al.\ 2015, \apj, 811, 139. doi:10.1088/0004-637X/811/2/139
\bibitem[Tian \& Chen(2018)]{Tian18} Tian, H. \& Chen, N.-H.\ 2018, \apj, 856, 34. doi:10.3847/1538-4357/aab15a
\bibitem[Veronig et al.(2010)]{Veronig10} Veronig, A.~M., Ryb{\'a}k, J., G{\"o}m{\"o}ry, P., et al.\ 2010, \apj, 719, 655. doi:10.1088/0004-637X/719/1/655
\bibitem[Warren et al.(2016)]{Warren16} Warren, H.~P., Reep, J.~W., Crump, N.~A., et al.\ 2016, \apj, 829, 35. doi:10.3847/0004-637X/829/1/35
\bibitem[Warmuth \& Mann(2020)]{Warmuth20} Warmuth, A. \& Mann, G.\ 2020, \aap, 644, A172. doi:10.1051/0004-6361/202039529
\bibitem[Weigelt(1977)]{Weigelt77} Weigelt, G.~P.\ 1977, Optics Communications, 21, 55. doi:10.1016/0030-4018(77)90077-3
\bibitem[Woods et al.(2017)]{Woods17} Woods, T.~N., Caspi, A., Chamberlin, P.~C., et al.\ 2017, \apj, 835, 122. doi:10.3847/1538-4357/835/2/122
\bibitem[Xiang et al.(2016)]{Xiang16} Xiang, Y.-. yuan ., Liu, Z., \& Jin, Z.-. yu .\ 2016, \na, 49, 8. doi:10.1016/j.newast.2016.05.002
\bibitem[Yan et al.(2018a)]{Yan18} Yan, X.~L., Yang, L.~H., Xue, Z.~K., et al.\ 2018a, \apjl, 853, L18. doi:10.3847/2041-8213/aaa6c2
\bibitem[Yan et al.(2018b)]{Yan18b} Yan, X.~L., Wang, J.~C., Pan, G.~M., et al.\ 2018b, \apj, 856, 79. doi:10.3847/1538-4357/aab153
\bibitem[Yan et al.(2020a)]{Yan20a} Yan, X., Liu, Z., Zhang, J., et al.\ 2020a, Science in China E: Technological Sciences, 63, 1656. doi:10.1007/s11431-019-1463-6
\bibitem[Yan et al.(2020b)]{Yan20b} Yan, X., Xue, Z., Cheng, X., et al.\ 2020b, \apj, 889, 106. doi:10.3847/1538-4357/ab61f3
\bibitem[Yan et al.(2021a)]{Yan21a} Yan, X., Wang, J., Guo, Q., et al.\ 2021a, \apj, 919, 34. doi:10.3847/1538-4357/ac116d
\bibitem[Yan et al.(2009)]{Yan09} Yan, Y., Zhang, J., Wang, W., et al.\ 2009, Earth Moon and Planets, 104, 97. doi:10.1007/s11038-008-9254-y
\bibitem[Yan et al.(2021b)]{Yan21} Yan, Y., Chen, Z., Wang, W., et al.\ 2021b, Frontiers in Astronomy and Space Sciences, 8, 20. doi:10.3389/fspas.2021.584043
\bibitem[Young et al.(2015)]{Young15} Young, P.~R., Tian, H., \& Jaeggli, S.\ 2015, \apj, 799, 218. doi:10.1088/0004-637X/799/2/218
\bibitem[Yu et al.(2020)]{Yu20} Yu, K., Li, Y., Ding, M.~D., et al.\ 2020, \apj, 896, 154. doi:10.3847/1538-4357/ab9014
\bibitem[Zarro et al.(1988)]{Zarro88} Zarro, D.~M., Canfield, R.~C., Strong, K.~T., et al.\ 1988, \apj, 324, 582. doi:10.1086/165919
\bibitem[Zhang et al.(2016a)]{Zhang16a} Zhang, Q.~M., Li, D., \& Ning, Z.~J.\ 2016a, \apj, 832, 65. doi:10.3847/0004-637X/832/1/65
\bibitem[Zhang et al.(2016b)]{Zhang16b} Zhang, Q.~M., Li, D., Ning, Z.~J., et al.\ 2016b, \apj, 827, 27. doi:10.3847/0004-637X/827/1/27
\bibitem[Zhang et al.(2019)]{Zhang19} Zhang, Q.~M., Li, D., \& Huang, Y.\ 2019, \apj, 870, 109. doi:10.3847/1538-4357/aaf4b7
\end{thebibliography}
\end{document}